\documentclass[showpacs,prl,twocolumn,floatfix,reprint]{revtex4}
\usepackage{graphicx,amssymb,amsmath}
\usepackage{esvect}

\begin{document}

\title{From one-way streets to percolation on random mixed graphs}

\author{Vincent Verbavatz}
\email{vincent.verbavatz@ipht.fr}
\affiliation{Institut de Physique Th\'{e}orique, CEA, CNRS-URA 2306, F-91191, 
Gif-sur-Yvette, France}

\author{Marc Barthelemy}
\email{marc.barthelemy@ipht.fr}
\affiliation{Institut de Physique Th\'{e}orique, CEA, CNRS-URA 2306, F-91191, 
Gif-sur-Yvette, France}
\affiliation{CAMS (CNRS/EHESS) 54 Boulevard Raspail, 75006 Paris, France}

\begin{abstract}

  In most studies, street networks are considered as undirected graphs
  while one-way streets and their effect on shortest paths are usually
  ignored. Here, we first study the empirical effect of one-way
  streets in about $140$ cities in the world. Their presence induces a
  detour that persists over a wide range of distances and
  characterized by a non-universal exponent. The effect of one-ways on
  the pattern of shortest paths is then twofold: they mitigate local
  traffic in certain areas but create bottlenecks elsewhere. This
  empirical study leads naturally to consider a mixed graph model of
  2d regular lattices with both undirected links and a diluted
  variable fraction $p$ of randomly directed links which mimics the
  presence of one-ways in a street network. We study the size of the
  strongly connected component (SCC) versus $p$ and demonstrate the
  existence of a threshold $p_c$ above which the SCC size is zero. We
  show numerically that this transition is non-trivial for lattices
  with degree less than $4$ and provide some analytical argument. We
  compute numerically the critical exponents for this transition and
  confirm previous results showing that they define a new universality
  class different from both the directed and standard
  percolation. Finally, we show that the transition on real-world
  graphs can be understood with random perturbations of regular
  lattices.  The impact of one-ways on the graph properties were
  already the subject of a few mathematical studies, and our results
  show that this problem has also interesting connections with
  percolation, a classical model in statistical physics.

  \end{abstract}


\maketitle

\section{Introduction}

In most countries a majority of individuals
commute by car \cite{Verbavatz:2019} and smart monitoring of traffic
in cities has become crucial for enhancing productivity while reducing
transport emissions \cite{Dodman:2009,Newman:2006}. Historically, a
simple and efficient way to manage traffic is by using dedicated
traffic codes, including the design of one-way streets
\cite{Lay:1992}. The first official attempt to create dedicated
one-way roads is said to date back to 1617 in London
\cite{Homer:2006}. The `No Entry' sign was officially adopted for
standardization at the League of Nations convention in Geneva in 1931
\cite{Lay:1992}. To this day, one-way streets are created in order to
smooth motor traffic in cities \cite{Stemley:1998}, to reduce driving time and
congestion, or to preserve specific neighborhoods
\cite{Venerandi:2004} from traffic. 

Mathematically, street networks
can be represented by graphs where the vertices are
intersections and the links road segments between consecutive
intersections. Almost all studies on street networks
\cite{Jiang:2004,Buhl:2006,Strano:2013,Xie:2007,Lammer:2006,Strano:2012,Crucitti:2006,Louf:2014,Barthelemy:2018,Boeing:2019}
describe street network as undirected graph but formally a network of both undirected links and one-way streets (represented by directed edges) is called a mixed graph \cite{Beck:2013}. Despite their relevance
for practical applications \cite{Roberts:1978}, there are very few
results available for directed street networks, except for the
following one: Robbins’ theorem \cite{Robbins:1939} states that it is
possible to choose a direction for each edge - called hereafter a
strong orientation - of an undirected graph $G$ turning it into a
directed graph that has a path from every vertex to every other
vertex, if and only if $G$ is connected and has no bridge (i.e. an
edge whose deletion increases the graph's number of connected
components). Robbins' seminal result can be extended to mixed graphs
\cite{Boesch:1980}, stating that if $G$ is a strongly connected mixed
graph, then any undirected edge of $G$ that is not a bridge may be
made directed without changing the connectivity of $G$. Hence, it is
possible to turn streets into one-ways as long as their removal does
not disconnect the whole street network. It is thus recursively
possible for any bridgeless network to be turned into a fully directed
graph. In most cities, it should then be possible to find a
street-orientation that keep the network strongly connected. This
theorem however does not say anything about how one-way streets modify
shortest paths. In this respect, very few results were obtained: for
the diameter for example, Chvatal and Thomassen \cite{Chvatal:1978}
proved that if the undirected graph has a diameter $d$, then there
exist a strong orientation with diameter less than the (best possible)
bound $2d+2d^2$, but that it is also a NP-hard problem to find. It is
interesting to note that for some applications, it is desirable to
find a strong orientation that is not efficient, i.e. doesn’t minimize
the diameter in order to discourage people from driving in certain
sections \cite{Roberts:1978}.

Here, we will first discuss some empirical results about the fraction
of one-way streets in cities and their effect on shortest paths. This will naturally
leads us to consider the problem of percolation in mixed graphs and the corresponding critical exponents that define a new universality class. We will then discuss the case
of real-world random graphs. 

\section{Empirical results}

Information about one-way streets in cities is available from
OpenStreetMap, an open source map of the world \cite{OSM}. We mined
this dataset with the open software OSMnX \cite{Boeing:2017} that
allowed us to extract directly the street network from $146$ cities
defined by their administrative boundaries. The graph analysis of real
networks was done with networkx \cite{Hagberg:2008} and the
theoretical analysis of regular lattices, computations of the
percolation threshold and of the critical exponents were done with the
C/C++ network analysis package igraph \cite{Csardi:2006}. The code is
available at \cite{GitVV}.

\subsection{Fraction of one-ways and detour index}

We define the fraction of one-way streets as $p=L_1/L(G)$
where $L_1$ is the total length of one-way streets and $L(G)$ the
total length of the network $G$ of size $N$. We observe that this
fraction ranges from very low values such as $8\%$ for the average of African cities
up to $31\%$ for the average of European ones. We show in Table
\ref{table:frac} the empirical value of $p$ in
  five different cities (compared to the SCC-percolation threshold in
  the corresponding graphs, see below).
\begin{table}
\begin{tabular}{|l|c|c|r|}
  \hline
  City & Country & One-way share ($\%$) & Threshold\\
  \hline
  Beijing & China & $37$  & $0.63(2)$\\
  \hline
    Casablanca & Morocco & $19$  & $0.73(2)$\\
  \hline
  Paris & France & $66$  & $0.78(2)$\\
  \hline
    New York City & USA & $55$  & $0.77(2)$\\
  \hline
    Buenos Aires & Argentina & $71$  & $0.78(2)$\\
  \hline
\end{tabular}
\caption{ Empirical fraction (in length) of one-way streets in
  five different cities compared to the SCC-percolation threshold in
  the corresponding graphs.  The percolation threshold is measured
  when the probability to have a giant cluster (connecting opposite
  sides) crosses $1/2$.}
\label{table:frac}
\end{table}

\begin{figure}
\centering
\includegraphics[width=0.5\textwidth]{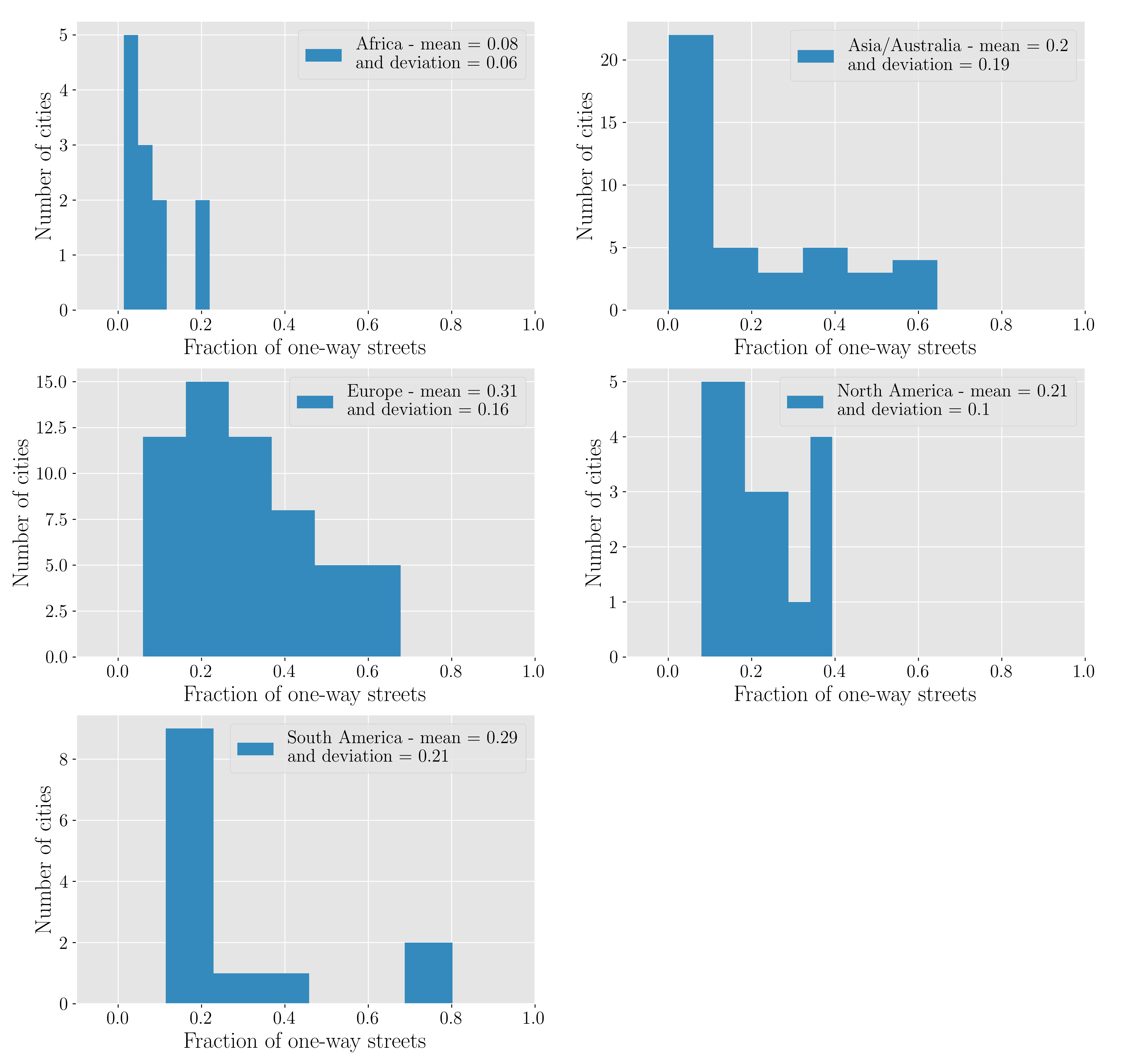}
\caption{Distribution of the fraction $p$ of one-way streets
  for the five continents (the fraction is defined as the total length
  of one-way streets over the total length of the network).}
\label{figSM0}
\end{figure}

We also show in Fig.~\ref{figSM0} the distribution of $p$ in different continents. In particular, we observe
that one-way streets are significantly more common in Europe than in the rest of
  the world.
The occurrence of one-way streets seems thus to be connected to more
complex street plans \cite{Boeing:2019}.

\begin{figure}
\centering
\includegraphics[width=0.45\textwidth]{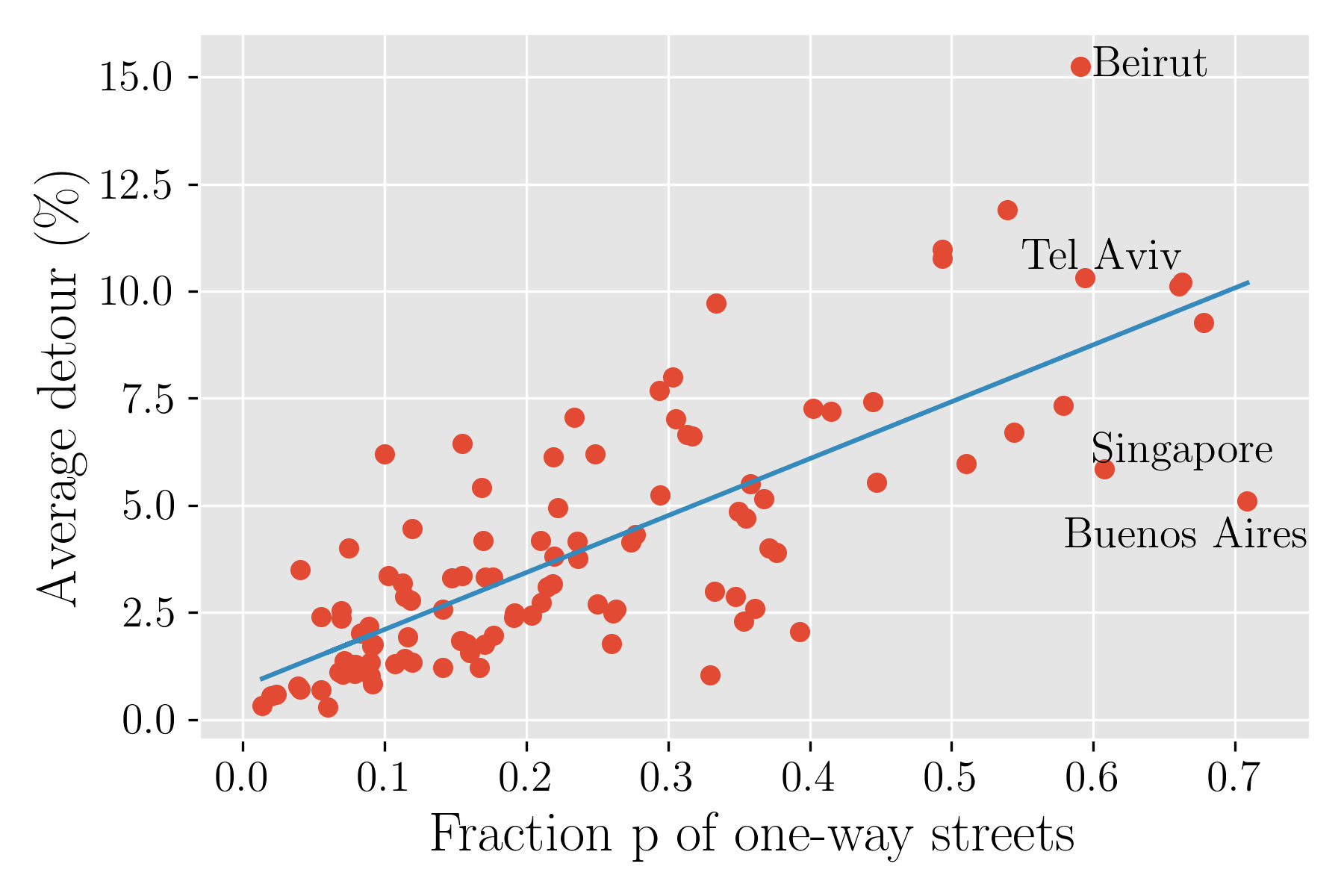}\\
\includegraphics[width=0.5\textwidth]{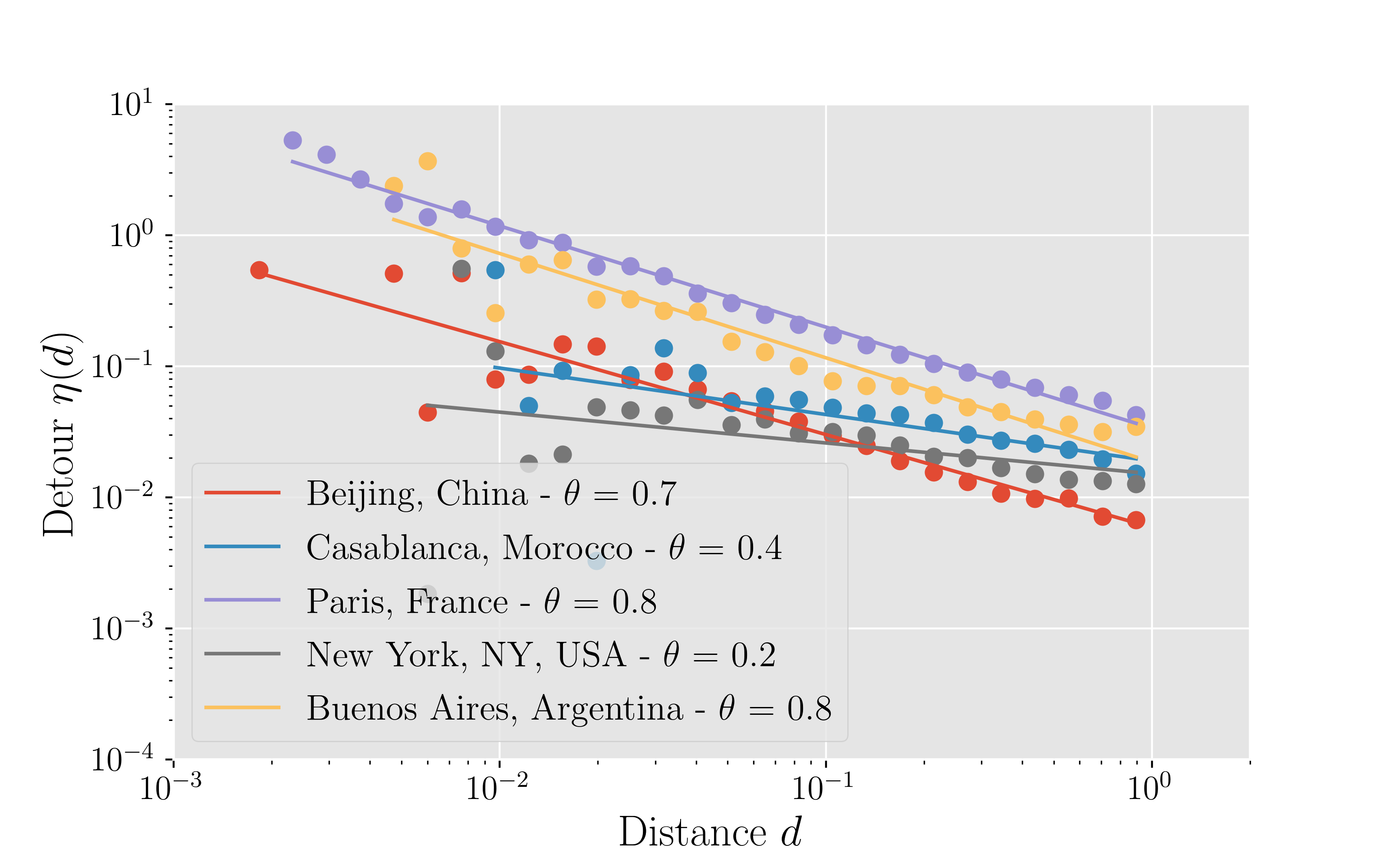}
\caption{(a) Distribution of the average detour ($\%$) as a function
  of fraction $p$ of one-way streets for 146 world cities ($R^2=0.59$). (b) For
  five selected cities in the world, we plot the average detour
  $\eta(d)$ due to one-way streets for a trip of distance $d$ as a
  function of $d$ (normalized by the maximum distance obtained for
  each city). The detour can be fitted by a power $\eta(d)\sim
  d^{-\theta}$. We
  find that $\theta$ differs from one city to another and ranges
  roughly from $0.2$ to $0.8$. In particular, small exponent values
  (such as in the case of NYC) might be correlated with the presence
  of very long one-way streets leading to a large detour even at very large
  spatial scales. We have $R^2=0.87$ for Beijing, $R^2=0.25$ for Casablanca, $R^2=0.99$ for Paris, $R^2=0.12$ for NYC and $R^2=0.90$ for Buenos Aires.}
\label{fig1}
\end{figure}

We denote by $d_G (i,j)$ the shortest path distance from node $i$ to
node $j$ on the undirected graph $G$ and $d_{\vv{G}}(i,j)$ the
corresponding quantity for the mixed graph denoted by $\vv{G}$ (when
one-ways are taken into account). The average detour due to one-ways
is then defined as
$\overline{\eta}=\frac{1}{N(N-1)} \sum_{(i,j)\in
  G}\frac{d_{\vv{G}}(i,j)}{d_G (i,j)} -1$. Figure~\ref{fig1}a shows
how the average detour increases with the fraction of one-way streets
$p$ in the dataset of world cities we use. We first observe that the
detour increases roughly linearly with the fraction of one-ways (a
power law fit gives an exponent of 0.8) and that most cities have an
average detour less than $10\%$. We also note that there is a large
dispersion of this detour for a given value of the one-way
fraction. For example, for $p\approx 0.6$ the detour varies from about
$6\%$ for Singapore up to $15\%$ for Beirut (and even $5\%$ for
$p=0.7$ for Buenos Aires), showing that the impact on shortest paths
depends strongly on the precise location of one-ways. Furthermore, we
can separate the impact of one-ways on various distances by defining
the detour profile given by
\begin{align}
\eta(d)=\frac{1}{N(N-1)}\sum_{(i,j)\;s.t.\;
  d_G(i,j)=d}\frac{d_{\vv{G}}(i,j)}{d_G(i,j)}-1
\end{align}
We observe for various cities on Fig.~\ref{fig1}b that $\eta(d)$
roughly decreases as a power law of the form $\eta(d)\sim d^{-\theta}$
demonstrating the impact of one-way streets even for large
distance (in this figure, the distance is normalized by its maximum
value for each city). In particular, we note that if on average the detour due to
one-way streets is of the order of 10\%, which seems small, detours at
short distances may be significantly higher (up to the order of
100\%). Also, even if $10\%$ is small at an individual level, this
has a non-neglibible effect in terms of time cost and congestion at the city scale when summed over all car
users. 

The exponent $\theta$ does not seem to be universal and ranges
between $0.2$ and $0.8$ for different cities. We note that we expect
in general $\theta\in [0,1]$ where the upper-bound $\theta=1$
corresponds to the case where one-way streets create a constant detour
in the directed network, implying $d_{\vv{G}}(i,j)=C +d_G (i,j)$ and
therefore $\eta(d)\sim 1/d$. The case $\theta=0$ corresponds to the
situation where the detour is proportional to the distance traveled:
$d_{\vv{G}}(i,j)\propto d_G (i,j)$ implying $\eta(d)\sim const$. In
any case, this slow decrease of $\eta(d)$ with $d$ signals the
long-range effect of one-ways on shortest paths.

\subsection{Betweenness centrality}

\begin{figure}
\centering
\includegraphics[width=0.5\textwidth]{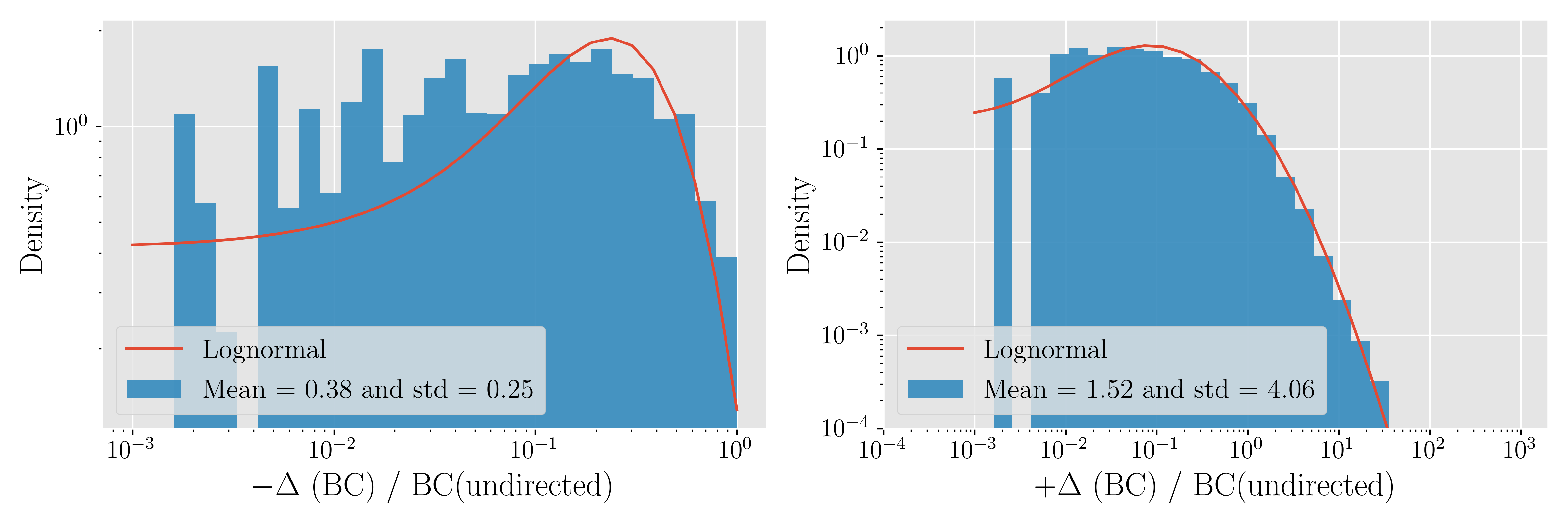}
\caption{Distribution of the relative variation $\Delta$ of the
  betweenness centrality (BC) due to one-way streets in the Parisian
  network for negative (the BC decreases) and positive values of $\Delta$
  (the BC increases with one-ways). Both distributions can be fitted
  by a lognormal and parameters are: $\mu=0.38$, $\sigma=0.25$ (negative values)
  and $\mu=1.52$ and $\sigma=4.06$. The distribution for positive values of $\Delta$ is
  much broader with large values of the relative variation of the BC
  demonstrating the creation of critical bottlenecks in the
  network. (a) For $53\%$ of the nodes, we have $\Delta<0$ which correspond
  to nodes having a smaller BC due to one-way streets. In this case,
  $27\%$ of the nodes have less than half the undirected BC and $3\%$
  less than $10\%$. (b) For $47\%$ of nodes, the BC is increased by
  one-way streets. For
  $31\%$ of these nodes, their BC doubled or more, and for $3\%$ it is
  ten times larger. }
\label{fig_bcentr}
\end{figure}

Cars have to follow the direction of links and consequently one-way
streets govern the spatial structure of traffic. The theoretical
question is then to understand what happens to the patterns of
shortest paths when we turn an undirected link into a one-way
street. This can for instance be measured by comparing the betweenness
centrality (BC) of nodes (see for example \cite{Kirkley:2018,Barthelemy:2018} and
references therein). We denote by $g_G (i)$ the BC of node $i$ on the
graph $G$ defined as
\begin{align}
  g_G(i)=\frac{1}{{\cal N}}\sum_{s\neq  t}\frac{\sigma_{st}(i)}{\sigma_{st}}
\end{align}
where $\sigma_{st}$ is the number of shortest paths from node $s$ to
node $t$ and $\sigma_{st}(i)$ the number of these shortest paths that
go through node $i$. The quantity ${\cal N}$ is a normalization that
we choose here  ${\cal N}=(N-1)(N-2)$. We denote by $g_{\vv{G}}(i)$
the BC of node $i$ when we include one-ways, and
we analyze the relative variation
$\Delta=(g_{\vv{G}}(i)-g_G(i))/g_G(i)$. In the case of Paris for
example, we find that $53\%$ of the nodes have a smaller BC
($\Delta<0$) due to one-way streets with $27\%$ of them having less
than half the undirected BC and $3\%$ less than $10\%$. For the other
$47\%$ with $\Delta>0$ the BC is increased, more than doubled for
$31\%$ of them and the BC is ten times higher in $3\%$ of cases. We
thus observe here the dual effect of one-way streets: certain nodes
are preserved and experience a reduced traffic while this
simultaneously create bottlenecks where the BC can be very
large. More generally, we observe (see Fig \ref{fig_bcentr}) that the
distribution of $\Delta$ is not symmetric (with a global average of $\sim
0.59$) and skewed towards positive
values indicating that the bottlenecks due to the deviated traffic can
be extremely busy.

\subsection{Strongly connected component}

\begin{figure}
    \includegraphics[width=0.5\textwidth]{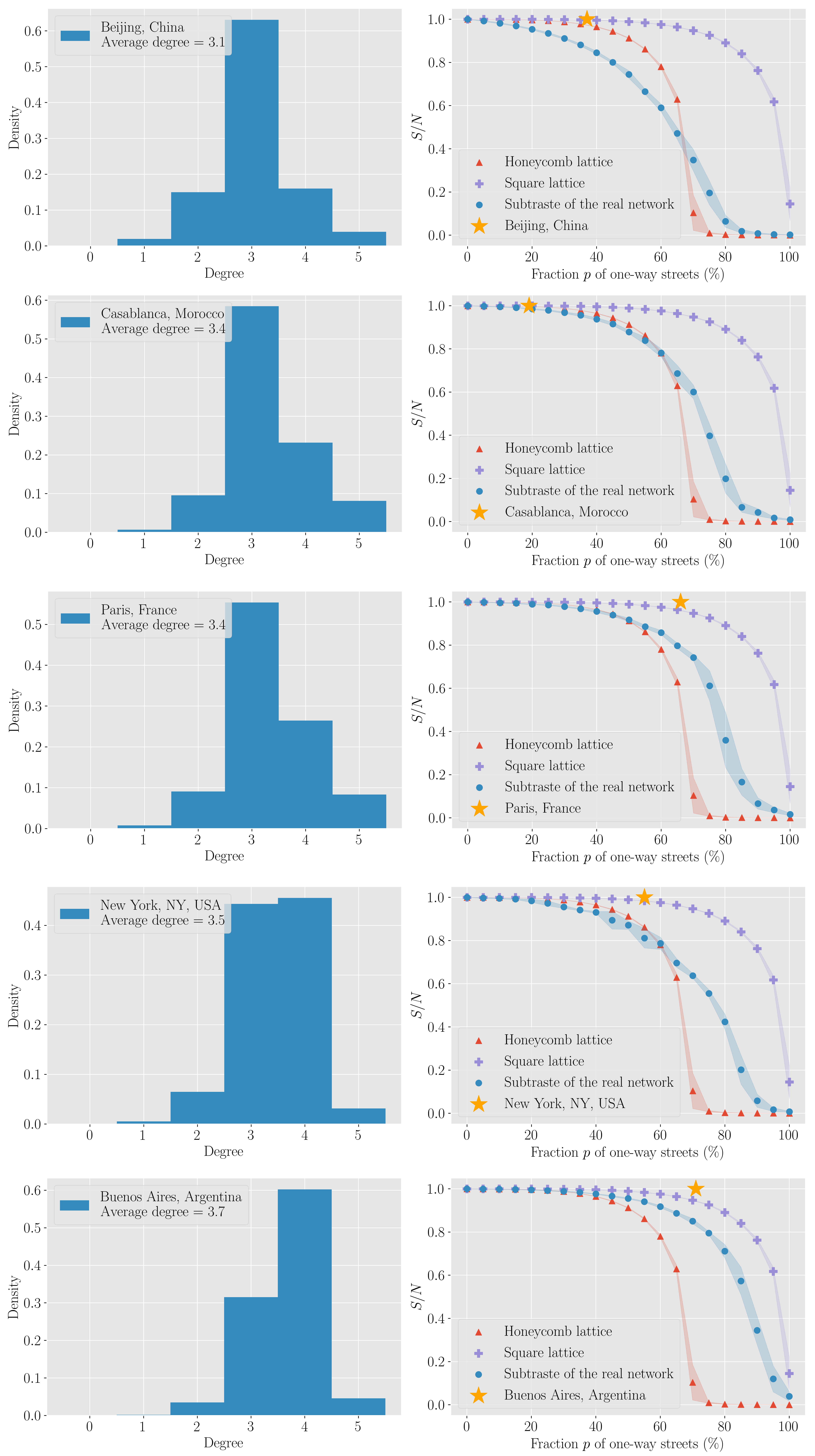}
    \caption{(Left column): The degree distribution of junctions for 5
      different cities from 5 different continents. The average degree
      for these cities is $\langle k\rangle\sim 3.4$ (Casablanca),
      $\sim 3.1$ (Beijing), $\sim 3.5$ (New York), $\sim 3.4$ (Paris),
      $\sim 3.7$ (Buenos Aires). The most common junction is a
      3-points fork in Casablanca, Paris and Beijing, while 4-points
      crossroads are more frequent in New York City and Buenos Aires.
      (Right column) The blue points are obtained by picking a
      fraction $p$ of streets in the underlying bidirectional
      structure of the city (that we call the substrate of the real
      network) and turning them into one-way streets. In that
      statistical process, bidirectional streets in the real world may
      be turned into one-way streets while one-way streets may be
      bidirectional. We then plot the largest strongly connected
      component size ($S$) in the total network normalized by the
      number $N$ of nodes as a function of $p$.  Results are obtained
      for 10 different disorder realizations.
    }
  \label{fig2}
\end{figure}

The strongly connected component (SCC) in the directed graph is the
set of nodes such that there is a directed path connecting any pairs
in it \cite{Roberts:1978}. We note that for a weakly connected graph
such as the street network, there is one SCC only. We first show (see
Fig.~\ref{fig2} left column) the distribution of degrees of nodes
(junctions) in five different cities in the world, whose fraction $p$
of one-way streets ranges from $19\%$ to $71\%$ (see Table~\ref{table:frac}). As we could anticipate, we note significant differences in the
degree distribution between old cities like Paris or Beijing where and
newer cities like New York City, where important areas are in the form
of a square grid. Except in the cases of Casablanca and Beijing,
one-way streets represent more than half of the total length of the
network. It is even more pronounced in the case of square-gridded
cities such as Manhattan where the percentage of one-ways is $69\%$
(with many east/west or north/south oriented avenues and streets)
which probably correspond to the need for decreasing congestion and
for simplifying the navigation in the city. For each of these cities,
we keep the underlying bidirectional structure of the graph (that we
call the substrate of the real network) and we vary the fraction $p$
of one-way streets from $0$ to $1$ by randomly turning a share $p$ of
streets into one-way streets (and $1-p$ is therefore the remaining
fraction of undirected links representing two-ways streets). In that
process, bidirectional streets in the real world may be turned into
one-way streets while one-way streets may be bidirectional.  Hence,
for each value of $p$, we randomly allocate one-way streets (with
random orientation) and compute the size $S$ of the strongly connected
component, normalized by the number $N$ of nodes. We construct many
realizations of this process allowing us to compute statistical
properties.

This measure of $S/N$ enables us to understand how many streets can be
randomly turned into one-way streets before parts of the city become
disconnected. We compare in Fig.~\ref{fig2} (right column) the
resulting curve for the same process on regular lattices of 3-point
junctions (honeycomb lattice) and 4-point junctions (square
lattice). For every city, we observe an abrupt percolation-like
transition for the SCC size when the fraction of random one-way
streets increases. We notice that for each city the real share
$p_{real}$ (represented by the star) of one-way streets is below the
transition threshold and that in general $({S/N})_{real}\approx 1$,
which means that - fortunately - cities are not disconnected in the
real life. This is expected for practical reasons and Robbins' theorem
\cite{Robbins:1939} states the existence of such a solution whatever
the fraction of directed links. We note, however, that this solution
is statistically not frequent and may be very far from the average of
${S/N}$ over all random configurations at share $p_{real}$.

\section{Percolation analysis}

\subsection{Percolation and digraphs. The model.}

These empirical results bring us to study in more depth this 
percolation-like transition observed for mixed graphs. We first note
that this problem is different from the rare results available for
digraphs (see for example
\cite{Luczak:1990,Newman:2001,Schwartz:2002,Doro:2001,Boguna:2005,Bianconi:2008}
and references therein). For example, similarly to the Erd\H{o}s-Renyi
transition \cite{Erdos:1960}, adding directed links to a digraph leads
to a transition for the strongly connected component
\cite{Luczak:1990}: for $M/N>1$, there is an infinite SCC ($M$ is the
number of directed arcs, and $N$ the number of nodes). The control
parameter is then the number of edges which are all directed. Other
studies generalized percolation in random fully directed – generally
uncorrelated – networks
\cite{Newman:2001,Schwartz:2002,Doro:2001,Boguna:2005} but whose
results cannot be directly applied to regular lattices due to the
strong degree correlations and the non-random nature of links. Our
model is also different from the well-known model of directed
percolation in statistical physics \cite{Obukhov:1980,Broadbent:1957}
where a preferred direction is chosen for all bonds on a regular
lattice and which defines a universality class different from usual
percolation.

This type of percolation model was introduced by Redner in a series of
papers \cite{Redner:1982a,Redner:1982b,Redner:1982c} as the random
resistor diode percolation, and was studied further in
\cite{Inui:1999,Janssen:2000,Zhou:2012,DeNoronha:2018}.  In the more
general version of this model defined on lattices, bonds can be
absent, be a resistor that can transmit an electrical current in
either direction along their length, or diodes that connect in one
direction only. The general phase diagram was discussed in
\cite{Redner:1982a,Redner:1982b} using real-space renormalization
arguments which predict fixed points associated with standard
percolation, directed percolation, and other new transitions. The
crossover between isotropic and directed percolation was further
studied in \cite{Inui:1999,Janssen:2000,Stenull:2001,Zhou:2012}. In
relation to the problem discussed here, Redner \cite{Redner:1982a} observed a `reverse
percolation' transition from a one-way connectivity in a given
direction to a two-way (isotropic) connectivity when connected paths
oriented opposite to the diode polarization begin to span the
lattice. This transition from a connected component to a strongly
connected component corresponds to what we observe here. 

The model discussed in this paper was previously considered in
\cite{DeNoronha:2018} where critical exponents are computed on
isotropically directed lattices where bonds can be either absent,
directed or undirected (in \cite{Hillebrand:2018} the authors
considered some properties in the critical case). The particular case where bonds are either
undirected or directed (but cannot be absent) is the specific case
that applies to road networks and that we will focus on. We recall
here the precise definition of this model. We consider a mixed graph $\vv{G}$ whose edges can be
either directed or undirected. As in the previous section, we denote
by $p$ the fraction of directed edges and the limits $p = 0$ and
$p= 1$ correspond then to the undirected and the fully directed graph,
respectively. We assume that the directed links have a random
direction without any bias (i.e. each direction has a probability
$1/2$). We vary the fraction $p$ and measure various quantities and we
will consider regular lattices such as the square and the honeycomb
lattices.

\subsection{Detour properties}

\begin{figure}
    \includegraphics[width=0.45\textwidth]{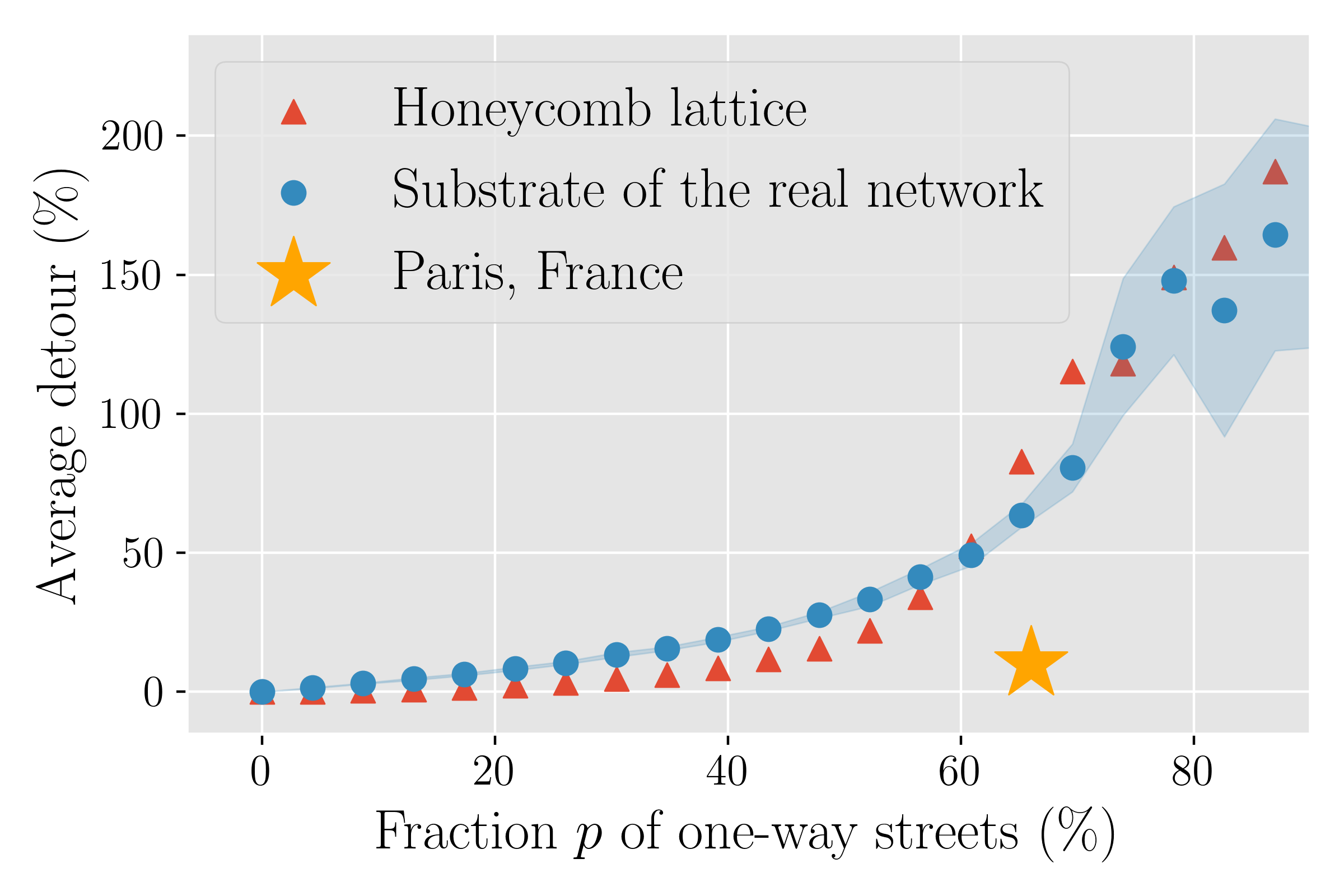}
    \caption{Average detour $\overline{\eta}$ as a function of the
      fraction $p$ of randomly chosen one-way streets in the city of
      Paris (France). In this statiscal process, the detour increases
      with the fraction $p$. We note, however, that the empirical
      detour in the real world (indicated by a star symbol) remains
      below the result expected from a random uniform distribution of
      one-way streets. This indicates that the actual choice of
      one-way streets in Paris is far from what would be obtained by a
      random choice of one-way streets and favors small detours. We
      compare these results to the obtained for a honeycomb lattice,
      whose degree distribution is close the Paris.}
\label{fig3}
\end{figure}

We will first consider the average detour on the honeycomb lattice and
observe that it increases with $p$ (Fig.~\ref{fig3} for Paris.
We also see in Fig.~\ref{fig3} that the real detour is below the
result obtained for a random distribution of one-way streets (similar
results are obtained for other cities). This demonstrates the
importance of the precise location of one-ways that can affect in very
different ways the shortest paths statistics.

\begin{figure}
\includegraphics[width=0.5\textwidth]{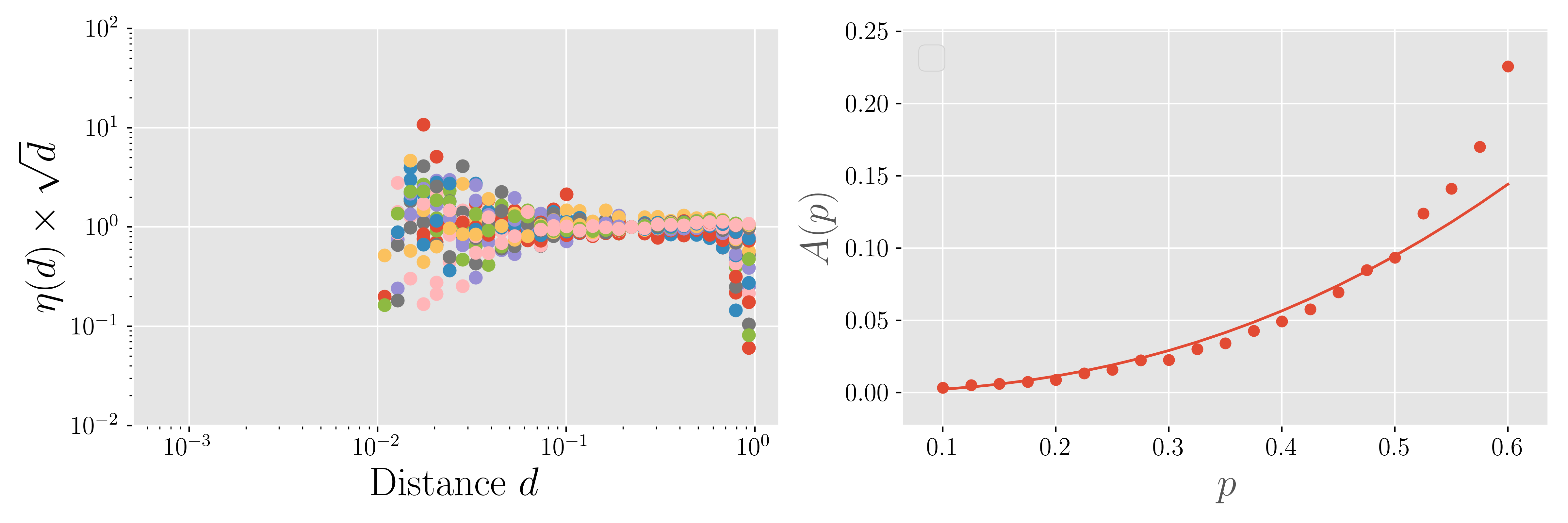}
\caption{ Average detour for the honeycomb lattice. (a) We show the quantity
  $\eta(d) \sqrt{d}$ versus d normalized by its value at $d=0.1$.  The curves
  collapse onto a single one independent from d and the observed
  discrepancies for $d$ close to $1$ and $d$ small come from finite-size
  effects. (b) The collapse suggest a form $\eta(d)= A(p)d^{-1/2}$, and a
  power law fit gives $A(p)\sim p^\gamma$ with $\gamma\approx 2.3$
  ($R^2=0.94$). We, however, observe discrepancies at large $p$.}
\label{figSM3}
\end{figure}

For the honeycomb lattice (Fig.~\ref{figSM3}), the average detour
$\eta(d)$ due to directed links for a trip of distance $d$ scales as a
power-law of $d$ with $\eta(d)\sim d^{-\theta}$ (the quantity $d$ is
here normalized by its maximum value). We find $\theta=0.5\pm 0.1$ as
shown in the data collapse of Fig.~\ref{figSM3}(a). More precisely, we
also show that the relation is of the form $\eta(d)=A(p) d^{-1/2}$
that remains valid for all $p$ and with $A(p)\sim p^{2.3}$ (see
Fig. \ref{figSM3}b). This result in $1/\sqrt{d}$ suggests the
possibility of an argument relying on the sum of random quantities
leading to $d_{\vv{G}}(i,j)-d_G (i,j)\sim \sqrt{d}$.

\subsection{Percolation threshold}

In the following, we will focus on the size of the SCC and related
properties. In order to distinguish the new transition from the usual
percolation we will use the term `SCC-percolation' when
needed. Similarly to classical percolation
\cite{Sykes:1964,Kesten:1982,Sahimi:1994,Callaway:2000,Christensen:2005,wiki,Stauffer:2018},
we denote by $P_\infty$ the probability to belong to the strongly
connected component and which will be the order parameter. We observe
numerically (over $1000$ runs) that both lattices exhibit a phase
transition (see Fig.~\ref{figSM1} and \ref{figSM2}) at a
percolation threshold $p_c$ above which the size of the SCC is
negligible. We determine the percolation threshold $p_c (L)$ for a
finite lattice of linear size $L$ using the method described in
\cite{Yonezawa:1989}. In order to determine the percolation threshold numerically, we define
the threshold $p_c (L)$ for a finite lattice of linear size $L$ as the
fraction of directed graphs for which the probability $P(L)$ to
observe a strongly connected cluster connecting two opposite sides of
the system is $0.5$ \cite{Yonezawa:1989}. In practice, we compute
$p_c (L)$ as the average threshold between the last time such that
$P(L)>0.5$ and the first time such that $P(L)<0.5$ when $p$
increases. Having the threshold
$p_c(L)$ for different sizes $L$, we use the classical ansatz \cite{Yonezawa:1989} 
\begin{align}
  p_c(L)=p_c(\infty)-A/L^\nu
\label{eq:nu}
\end{align}
where $\nu$ is the exponent that describes the divergence of the correlation length
$\xi\sim |p-p_c |^{-\nu}$. Using this method, we find for the
honeycomb lattice $p_c= 0.6935 \pm 0.0005$ and
$p_c= 0.998 \pm 0.002$ for the square lattice (see Fig.~\ref{figSM1} and Fig.~\ref{figSM2}). For honeycomb lattices
we thus observe a threshold $p_c<1$ while for the square lattice we have
$p_c=1$. This means here that for a degree equal or larger than $4$,
the number of different paths between any pair of points is large
enough so that the SCC is always large. In contrast, for the honeycomb
lattice with a degree $k=3$, some nodes can more easily constitute
`blocking points' with one-way streets ending at it (see below for a
more detailed argument). Interestingly enough, real street networks
have an average degree between $3$ and $4$ implying a non-trivial
threshold and the corresponding curve to lie between those for the two
lattices. The scaling ansatz also gives the
value $\nu= 1.1 \pm 0.2$ (and the same value for the square
lattice) which is slightly different from the isotropic percolation value $4/3$.

For this model, de Noronha et al.~\cite{DeNoronha:2018} proposed a
conjecture for computing the percolation threshold which is based
on the idea that it is governed by the probability that the
nearest-neighbor can be reached from a given site. Using duality
arguments, this conjecture can be proven to be exact for the square,
triangular, and honeycomb lattices \cite{DeNoronha:2018}. For the model where
bonds are either undirected or directed (but not absent), this
conjecture reads
\begin{align}
  p_c=2(1-p_c^0)
  \label{eq:con}
\end{align}
where $p_c^0$ is the corresponding threshold for the usual percolation on the
lattice. For the honeycomb lattice, $p_c^0=1-2\sin\pi/18$ which
implies $p_c=4\sin\pi/18\approx 0.6926...$ in agreement with our
numerical estimate. This conjecture was tested on both the honeycomb
and square lattices only and we tested it on real-world random graphs
for different cities. We show the results in Table~\ref{table:pc}.
\begin{table}
\begin{tabular}{|l|c|c|r|}
  \hline
  City & $p_c(SCC)$  & $p_c=1-\frac{1}{2}p_c(SCC)$ & $p_c$ (measured)\\
  \hline
Beijing & $0.63$ & $0.685$ & $0.67(3)$ \\
  \hline
Casablanca & $0.73$ & $0.635$   & $0.62(3)$  \\
  \hline
Paris & $0.78$ & $0.61$   & $0.57(3)$  \\
  \hline
NYC & $0.77$ & $0.615$   & $0.57(3)$  \\
  \hline
Buenos Aires & $0.88$ & $0.56$   & $0.52(3)$  \\
  \hline
\end{tabular}
\caption{We show here the SCC percolation threshold for different
  cities ($p_c(SCC)$), the percolation threshold predicted using the
  conjecture Eq.~\ref{eq:con} proposed in \cite{DeNoronha:2018}, and
  the measured threshold.}
\label{table:pc}
\end{table}
We observe that there is a good agreement between the value predicted
by the conjecture Eq.~\ref{eq:con} and our direct measure for
different cities: the conjecture seems to be correct for these random
graphs (within our error bars). 

\begin{figure}
\centering
  \includegraphics[scale=0.5]{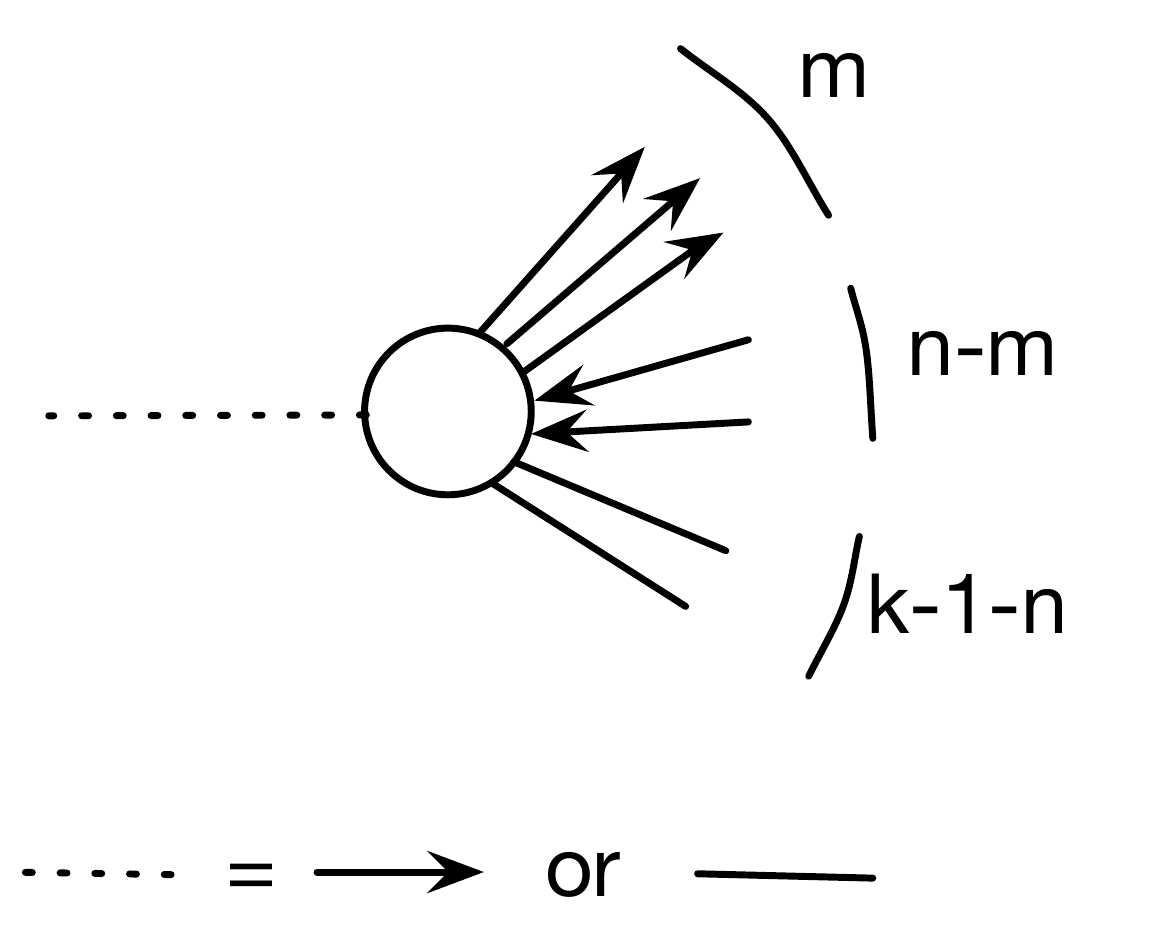}
  \caption{Notations: a node of degree $k$ has an incoming link and
    $k-1$ outgoing links. Among those, we have $m$ outgoing links,
    $n-m$ incoming links and $k-1-n$ bidirectional edges.}
  \label{fig:calcul}
\end{figure}

This conjecture shows that once $p_c^0$ is smaller than $1/2$, there is
no transition. For a regular lattice of degree $k$ (which is
$k=2d$ for a hypercubic lattice in dimension $d$), we can then ask
what is the value of $k$ above which there is no transition
anymore. The percolation threshold is obviously an increasing function of the
lattice degree $k$, as it is easier to find a strongly connected
component on graphs with more neighbors, and there seems to be no
transition for lattices with average degree larger than $4$.  It is
easy to show that $p_c=0$ for the one-dimensional lattice (which
corresponds to a regular lattice with degree $k=2$). We propose the
following approximation in order to understand how the threshold
varies with the degree $k$ in a regular lattice. We adapt to our case
the argument proposed in \cite{Schwartz:2002}: we assume that a node
has an incoming link and we compute its average outdegree
$\langle k_o\rangle$ (which varies from $0$ to $k-1$, we do not take
into account the incoming link here). The notations used are defined in Fig.~\ref{fig:calcul}.
The probability of having the links defined by $(n,m)$ is given by
\begin{align}
  p_{nm}=\left(\frac{p}{2}\right)^n (1-p)^{k-1-n}
\end{align}
We take into account that the incoming link can be either undirected (with
probability $1-p$) or directed and incoming with probability $p/2$
leading to a prefactor $p/2+1-p$. The outdegree
for the configuration defined by $n$ and $j$ is $k-1-n+m$. Considering
also the combinatorial factors, we obtain
\begin{align}
  \langle k_o\rangle
 &=\sum_{n=0}^{k-1}\left(\frac{p}{2}\right)^n(1-p)^{k-1-n} 
\sum_{m=0}^{n}\binom{k-1}{m}\binom{k-1-m}{n-m}  \\ 
  &\times (k-1-n+m)\left[\frac{p}{2}+1-p\right]
\end{align}
These sums can easily  be computed and we find
\begin{align}
  \langle k_o\rangle =\left(1-\frac{p}{2}\right)^2(k-1)
\end{align}
The percolation condition is
then $\langle k_o\rangle \geq 1$ which means that a directed path can
go through this node which is a necessary condition for belonging to
the SCC. Writing $\langle k_o\rangle=1$ then gives the percolation threshold
\begin{align}
  p_c(k)=2\left(1-\frac{1}{\sqrt{k-1}}\right)
\label{eq:pc}
\end{align}
which is valid in the interval $[2,5]$. This approximate formula gives the exact result $p_c (k=2)=0$ and
$p_c (k\geq 5)=1$. The latter is obviously an approximation but it is in
agreement, at least qualitatively with our numerical results. It
however overestimates - as expected for a necessary but not sufficient
condition - the degree above which $p_c=1$, and it would be
interesting to find how to modify this argument in order to recover
the numerical result $p_c (k=4)=1.0$.

\subsection{Critical exponent estimates: a new universality class}

\begin{figure}
\includegraphics[width=0.5\textwidth]{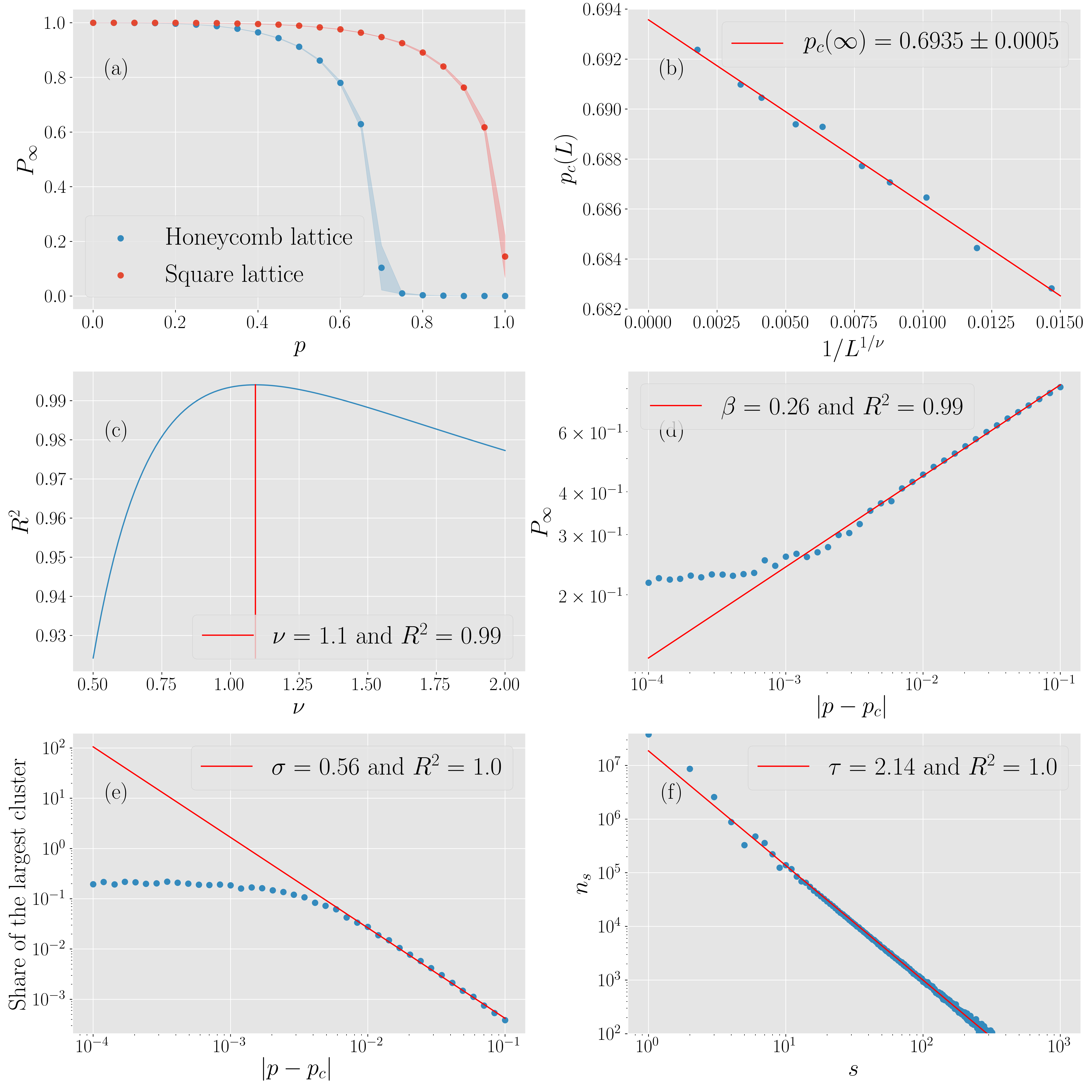}
\caption{SCC-percolation
  transition for the mixed honeycomb lattice and the
  calculation of critical exponents (too close to
  criticality, finite-size effects become important when the
  correlation length is of order the system size which reduces the
  range over which the fit can be made). (a) The probability to belong to the infinite cluster $P_\infty$
  drops dramatically when the fraction $p$ of one-way streets is close
  to $0.69$ in the honeycomb lattice and $1$ in the square
  lattice.  (b) Calculation of $p_c=0.6935\pm
  0.0005$. (c) The regression of the finite-size
  percolation threshold as a  function of $L$ gives the exponent $\nu=
  1.1 \pm 0.2$. (d) Below criticality, the behavior of $P_\infty$ with $|p-p_c |$
  gives the exponent $\beta=0.26\pm 0.02$. (e) Above criticality, the maximal
  normalized cluster size scales as $s_{max}\sim |p-p_c|^\sigma$ and
  we  find $\sigma= 0.56 \pm 0.05$. (f) At criticality, the
  number of clusters of sizes $s$ scales as $n_s\sim s^{-\tau}$ and we
  find $\tau= 2.14 \pm 0.05$.}
    \label{figSM1}
\end{figure}

\begin{figure}
\centering
\includegraphics[width=0.5\textwidth]{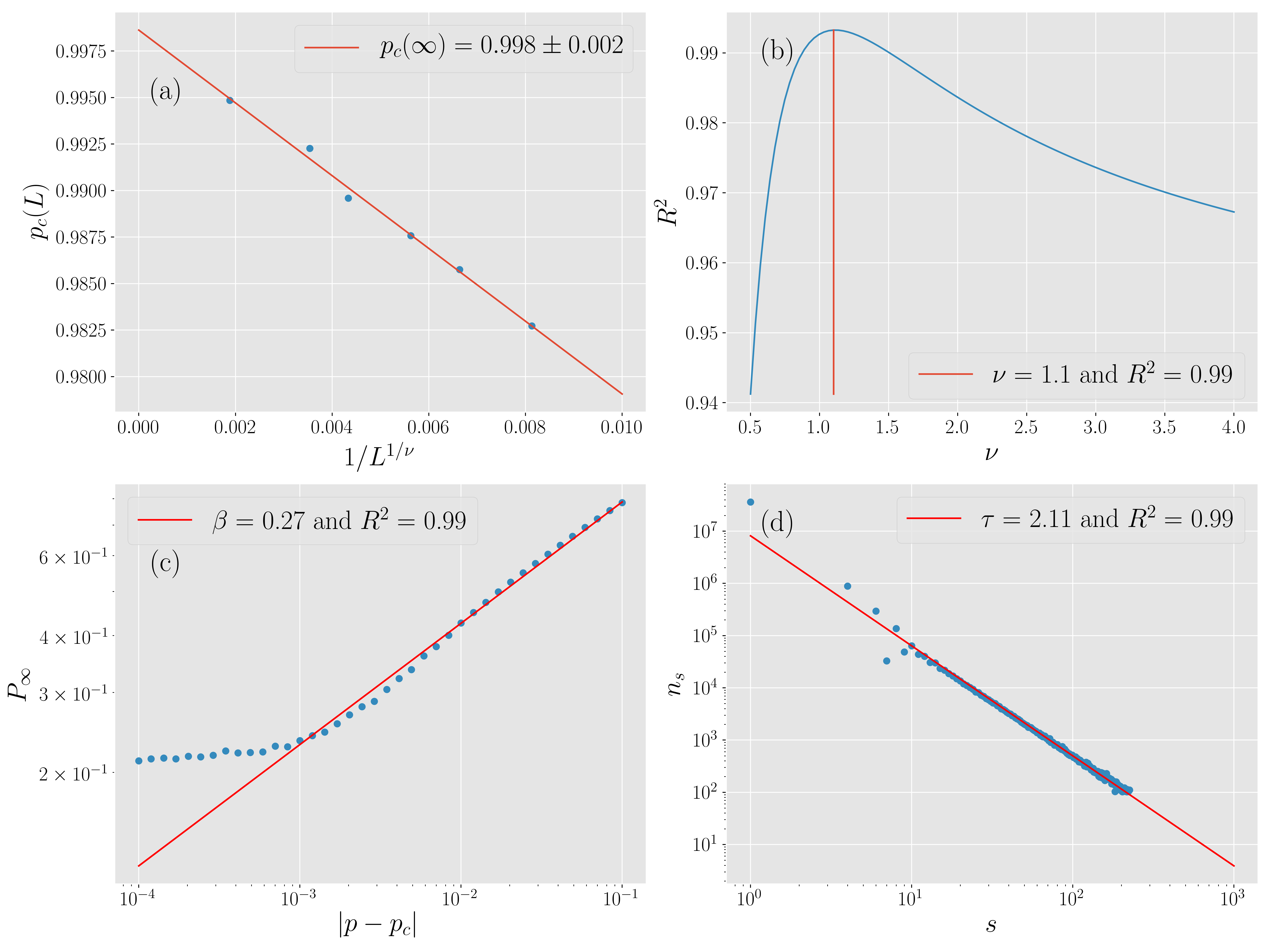}
\caption{(a) The percolation threshold for an infinite
  square lattice is calculated as an extrapolation for various
  finite-size lattices of side size ranging from $L= 100$ to
  $L= 1000$. We find $p_c = 0.998\pm 0.002$. (b) The regression of the
  finite-size percolation threshold as a function of the linear size
  also gives the critical exponent $\nu$, and we obtain
  $\nu= 1.1\pm 0.2$. (c) Below criticality, the behavior of $P_\infty$
  with $|p-p_c |$ gives the exponent $\beta=0.26$. (d) At criticality,
  the number of clusters of sizes $s$ scales as a power-law of the size
  with critical exponent $\tau$ and we find $\tau= 2.14\pm 0.05$.}
\label{figSM2}
\end{figure}

The critical exponents for this model were already estimated in
\cite{DeNoronha:2018} and we determine them independently for
 both the honeycomb (Fig.~\ref{figSM1}) and the
square lattices (Fig.~\ref{figSM2}). In particular, in
\cite{DeNoronha:2018} it is assumed that the exponent $\nu$ is the same as
in isotropic percolation and given by $\nu=4/3$. We replaced here this
assumption by the scaling ansatz Eq.~\ref{eq:nu} form for the percolation
threshold. 

Below the percolation threshold, the order parameter scales as
$P_\infty\sim |p-p_c |^\beta$ and a direct fit (Fig.~\ref{figSM1}d) gives
$\beta=0.26\pm 0.02$ ($0.27\pm 0.02$ for the square). Above the
percolation threshold, the maximal cluster size scales as
$s_{\max}\sim |p-p_c |^\sigma$ and at the threshold exactly, the
probability $n_s$ to belong to a cluster of size $s$ scales as
$n_s\sim s^{-\tau}$. These classical exponents take here the following
values (Fig.~\ref{figSM1}): $\tau= 2.14 \pm 0.05$ ($2.11 \pm 0.05$ for the square lattice)
and $\sigma= 0.56 \pm 0.05$ (the exponent $\sigma$ is not defined for
the square lattice where $p_c=1$). We note here that too close to criticality however, finite-size
effects become important when the correlation length is of order the
system size which reduces the range over which the fit can be made.
For the square lattice, we obtain the exponents in a similar way
(Fig.~\ref{figSM2}).

We note that these exponents satisfy the
hyper-scaling relations \cite{Christensen:2005} $\tau=d\sigma\nu+1$
and $\beta=(\tau-2)/\sigma$ (where the dimension is here $d=2$), which
is expected as these relations are independent from the fact that
links are oriented or not. From the classical relations
$d_f=d/(\tau-1)$ we get for the fractal dimension of the SCC at the
threshold the value $d_f=1.75\pm 0.08$.

We summarize these results in Table~\ref{table:expo}. We observe that the exponents are very different from the ones
obtained for the percolation on regular undirected lattices or for the
directed percolation, in agreement with the results obtained in
\cite{DeNoronha:2018} and pointing to a new universality class in
contrast with the analysis presented in \cite{Stenull:2001,Zhou:2012}
that showed that this model is in the same universality class as
standard percolation. There are however some numerical discrepancies (for $\nu$,
$\sigma$, and $d_f$) between our results and those of
\cite{DeNoronha:2018} and further work would be needed for a precise
determination of the exponents. 
\begin{table}
\begin{tabular}{|l|c|c|c|r|}
  \hline
  Critical & 2d  & 2d directed & Results &  This study\\
   exponent &    percolation   & percolation & of \cite{DeNoronha:2018}  & \\
  \hline
  $\nu$ & $4/3$ & $1.73$ (parallel) & $4/3$ & $1.1\pm 0.2$ \\
             &             & $1.09$ (perp.)   & & \\
  \hline
  $\beta$ & $0.14$ & $0.28$   & $0.27\pm 0.01$ & $0.26\pm 0.02$  \\
  \hline
  $\sigma$ & $0.40$ & $0.31$ &  $0.41\pm 0.01$ & $0.56\pm 0.05$ \\
    \hline
  $d_f$ & $1.90$ & $1.84$      &  $1.80\pm 0.01$ & $1.75\pm 0.08$ \\
    \hline
   $\tau$ & $2.05$ & $1.46$  & $2.12\pm 0.08$ & $2.14\pm 0.05$ \\
  \hline
\end{tabular}
\caption{Critical exponents for standard percolation
  \cite{wiki,Jensen:1999} compared to directed percolation
  \cite{Deng:2018}, the results obtained in \cite{DeNoronha:2018}, and
  our results for SCC-percolation on mixed graphs.}
\label{table:expo}
\end{table}

\section{Understanding the transition in disordered real-world
  networks}

\begin{figure}
\includegraphics[width=0.5\textwidth]{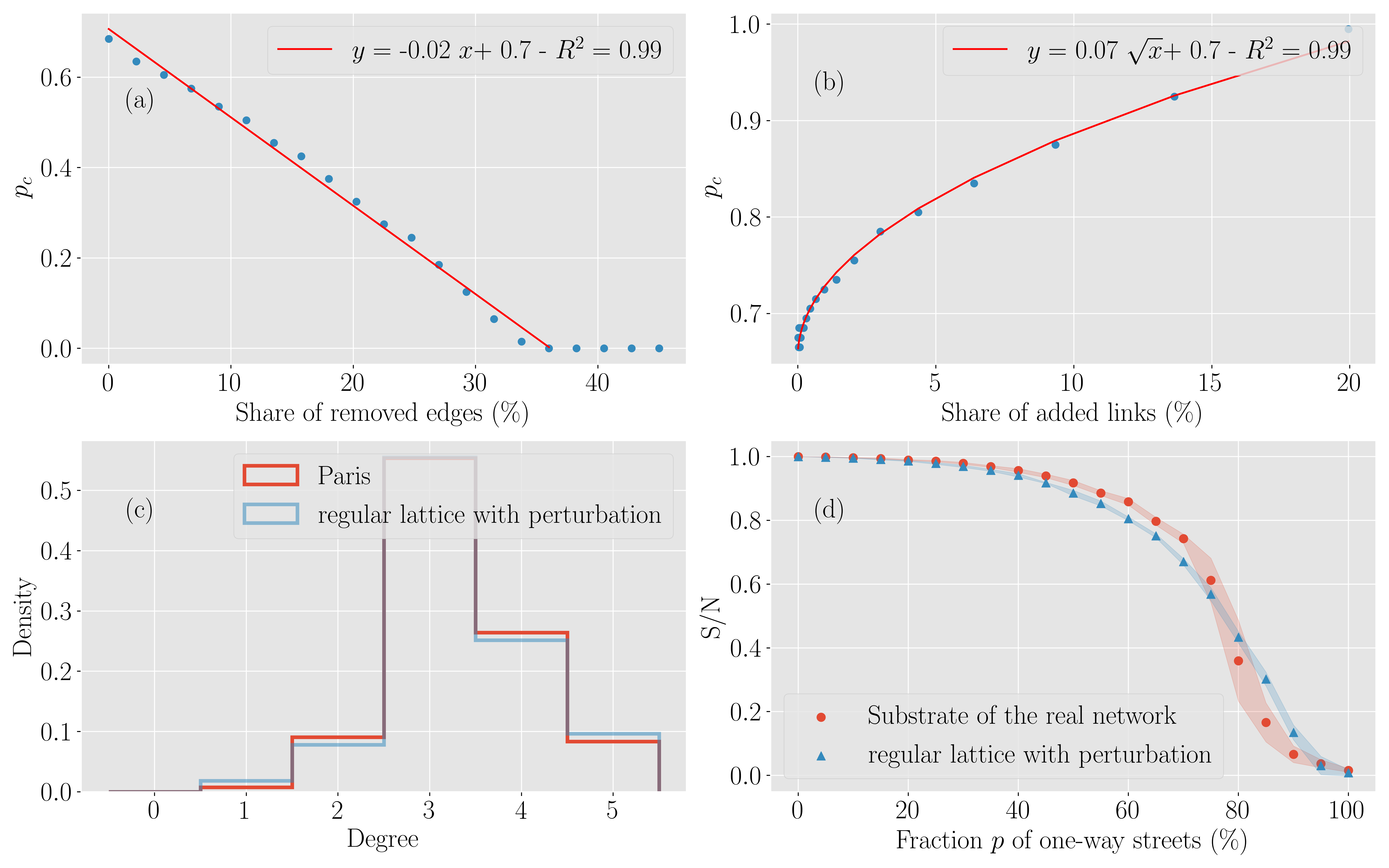}
\caption{(a) The SCC-percolation threshold decreases linearly
  with the share of edges
  removed from the honeycomb lattice. When the fraction of removed
  links is about $35\%$, the giant component of the undirected
  honeycomb lattice breaks down and the SCC-percolation threshold is
  0. (b) The SCC-percolation threshold increases with the number of
  edges added to the honeycomb lattice. The behavior is here well
  fitted by a square root function. (c) Starting from a regular square
  lattice, we construct various random planar graphs by both addition
  and removal of edges until the distribution of degrees is close to
  Paris. (d) On average, we recover the SCC-percolation transition of the Paris real network.}
  \label{fig4}
\end{figure}

Real-life street networks differ from the theoretical square and
honeycomb lattices. In particular, the degree distribution of vertices
(junctions) in city networks can exhibit different shapes (see
Fig.~\ref{fig2} left), either being centered around 3-point junctions
- like in Beijing - and hence closer to the honeycomb lattice, or
being centered around 4-point junctions – as in Buenos Aires for
instance - and closer to the square lattice, or being a combination of
both like in New York City. In order to test the effect of disorder on
the percolation behavior, we build various graphs starting from
regular lattices, and add or remove randomly edges. Removing links
from the honeycomb lattice shifts the SCC-percolation threshold
towards lower values in a linear way (Fig.~\ref{fig4}a) while the
average degree $\langle k\rangle$ drops below $3$. When the fraction
of removed links is about $35\%$ which corresponds to the standard
bond percolation threshold of the regular undirected honeycomb lattice
(the exact value is $2\sin\pi/18$ \cite{Sykes:1964}), the giant
component vanishes even without directed links (an obvious necessary
condition for having a SCC is indeed the existence of a weakly
connected giant component). On the contrary, adding random edges to
this graph increases the percolation threshold until they are too many
edges in the system and the transition does not occur anymore, as
there is always a directed path connecting any pair of nodes
(Fig.~\ref{fig4}b).

\begin{figure}
\includegraphics[width=0.5\textwidth]{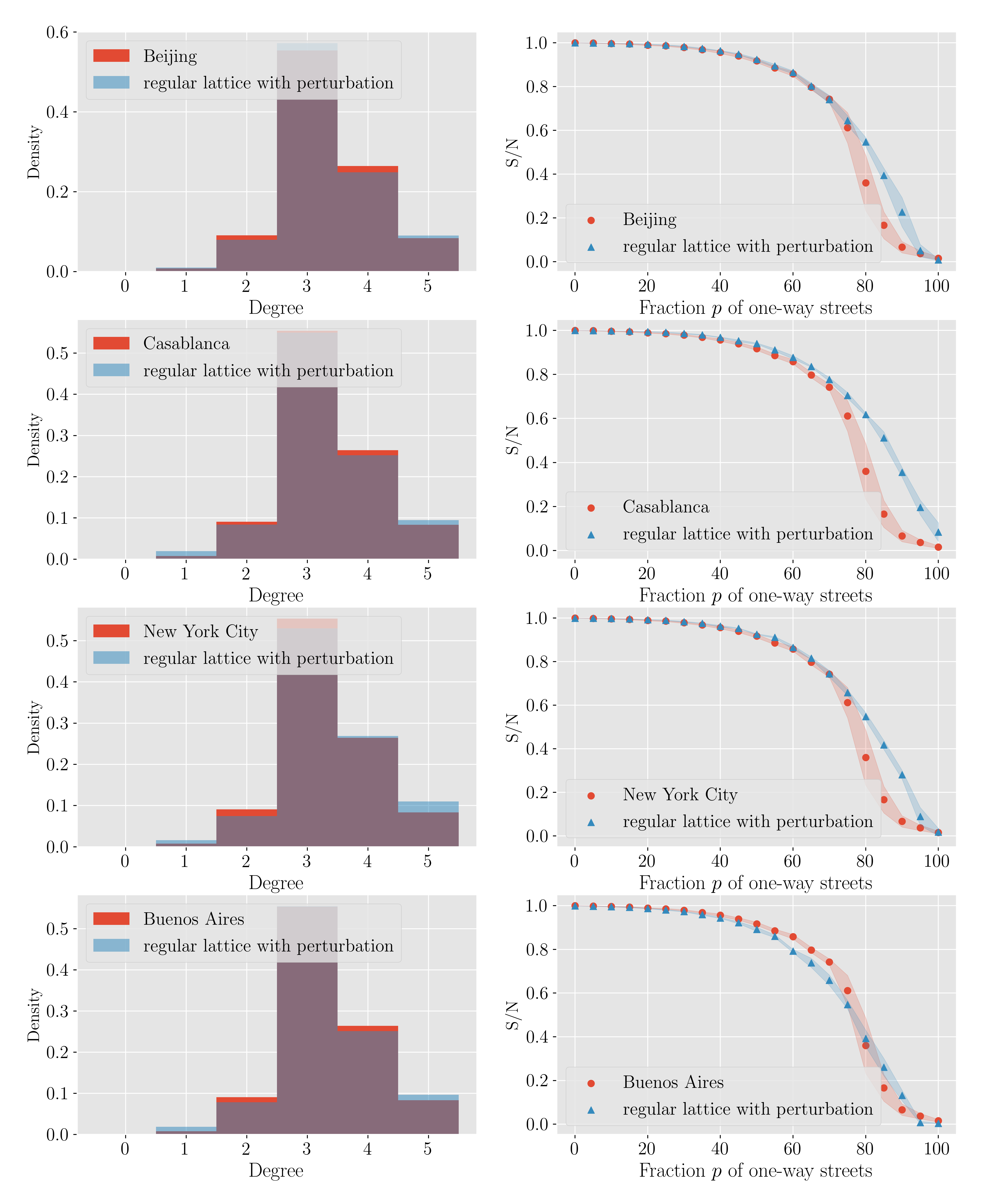}
\caption{In four cities, starting from a regular square
  lattice, we construct various random planar graphs by both addition
  and removal of edges until the distribution of degrees is close to
  the real one (left panel). On average, we recover the
  SCC-percolation transition of the corresponding real network (right
  panel).}
\label{figSM5}
\end{figure}

As observed above (Fig.~\ref{fig2} right column), underlying graphs of
real-world networks exhibit different non-trivial SCC-percolation
behaviors that result from the disorder in their structure. We model
these graphs by removal and addition of links in the
regular graph. There are several different ways of generating a random planar graph
whose distribution of degrees is close a given distribution. To
approximate the degree distribution of real world cities, we use the following heuristic
algorithm: starting from a regular square lattice, we delete a certain share
$\alpha_4$ of links for which at least one of the endpoints has degree
4. We then do the same operation by removing a certain share of links
$\alpha_3$ for which at least one of the endpoints has degree 3, then
2. Finally, we add a share of links $\beta_4$ between nodes of degree
4 and other nodes. We then adjust step by step the parameters
$\alpha_1$, $\alpha_2$, $\alpha_3$, $\alpha_4$ and $\beta_4$ until we
find a distribution of degrees that is reasonably close to the real
one. We test this model on the case of Paris (France)
and we construct a random mixed graph whose distribution of degrees is
close to the real one: starting from a
regular square lattice, we construct various random planar graphs by
both addition and removal of edges until the distribution of degrees
is close to the empirical one (for Paris here). With this theoretical
network, we are able to recover the observed percolation transition of
the underlying network of Paris (Fig.~\ref{fig4}c and d) not to be
confused with the actual choice of one-way streets in Paris, which was
proven to be statistically unlikely. We retrieve the transition both
at the level of the percolation threshold and the shape of the
function (see Fig.~\ref{figSM5} for other cities).

These results suggest that the degree distribution is actually the main determinant for the
percolation behavior on these real-world graphs. It is important to
note that for percolation, bonds are drawn at random, while as noted
above, there are correlations between one-way streets locations in
real configurations and the degree distribution is not the only
determinant in this case.

\section{Discussion}

One-way streets in large cities are of fundamental importance for
controlling car traffic with dramatic effects on neighborhoods in
terms of pollution and noise. Urban planners have achieved to increase
the number of one-way streets in cities while preserving a giant
strongly connected component, as ensured by Robbin’s theorem: even if
it is a very hard task to do from scratch, adding one-ways by
preserving the strong orientation is a working strategy. How to locate
one-way streets and their effect on the graph structure were already
the subject of a few mathematical studies in graph theory, and we show
here that this problem has in addition interesting connections with
statistical physics. In particular, this problem naturally leads to a
new percolation-like model which belongs to a new universality
class. Understanding better this transition on both regular
lattices and disordered graphs represents certainly a challenge for
theoretical physicists, and might also shed light on the effects of
one-way streets in our cities.

{\it Acknowledgements} - This paper is dedicated to the memory of Pierre Rosenstiehl who
recently passed away. We thank Geoff Boeing for his invaluable help
for using OSMnX and Sid Redner for useful discussions about the random
resistor-diode network. MB thanks Edouard Br\'ezin for his original suggestion
to look at this problem and Fabien Pfaender for discussions at an
early stage of this work. This material is based upon work supported
by the Complex Systems Institute of Paris Ile-de-France (ISC-PIF). VV
thanks the Ecole nationale des Ponts et Chauss\'ees for financial
support.

\bibliographystyle{prsty}

\end{document}